# Kagome bands and magnetism in MoTe$_{2-x}$ kagome monolayers

Jiaqi Dai[1,2,†], Zhongqin Zhang[1,2,†], Zemin Pan[3,4], Cong Wang[1,2], Chendong Zhang[3,4]*, Zhihai Cheng[1,2] and Wei Ji[1,2,*]

[1]*Beijing Key Laboratory of Optoelectronic Functional Materials & Micro-Nano Devices, School of Physics, Renmin University of China, Beijing 100872, China*
[2]*Key Laboratory of Quantum State Construction and Manipulation (Ministry of Education), Renmin University of China, Beijing 100872, China*
[3]*School of Physics and Technology, Wuhan University, Wuhan 430072, China*
[4]*Wuhan Institute of Quantum Technology, Wuhan 430206, China*

*Emails: wji@ruc.edu.cn (W.J.); cdzhang@whu.edu.cn (C.Z.)

**ABSTRACT:**

Kagome lattices facilitate various quantum phases, yet in bulk materials, their kagome flat-bands often interact with bulk bands, suppressing kagome electronic characteristics for hosting these phases. Here, we use density-functional-theory calculations to predict the geometric and electronic structures, as well as the topological and magnetic properties, of a series of MoTe$_{2-x}$ kagome monolayers formed by mirror-twin-boundary (MTB) loops. We analyze ten MTB-loop configurations of varying sizes and arrangements to assess their impact on various properties. Within the intrinsic bandgap of MoTe$_2$, we identify two sets of kagome bands, primarily originating from in-plane and out-of-plane Mo $d$-orbitals at MTB-loop edges and -vertices, respectively. Three configurations exhibit superior stability, while three others show comparable stability. Among these, four display bandgaps and potentially non-zero $Z_2$ topological invariants, suggesting possible topological phases, while the remaining two are metallic and feature Stoner magnetization. These findings guide the design of kagome-based two-dimensional materials with tunable electronic, topological, and magnetic properties.

**Introduction**

Kagome materials, characterized by their lattice of corner-sharing triangles, have attracted significant attention for their unique electronic properties, such as electronic flat-bands, Dirac cones, and von Hove singularities[1,2]. These features, combined with strong electron correlation, enable diverse quantum phases such as magnetism[3,4], superconductivity[5,6], and charge density waves (CDW)[7,8], making kagome lattices prime candidates for studying correlated and topological electronic phenomena[1,2,4,9]. While bulk kagome materials exhibit intriguing properties[1,2,5,10], such as the magnetic Weyl semimetal phase in $Co_3Sn_2S_2$[4,11–13] and superconductivity in $AV_3Sb_5$ (A=alkali metals)[5,14], their electronic structures are often complicated by coupling with capping layers[4,15–17]. This coupling causes characteristic kagome bands, like flat bands, to mix up with other bulk bands[4,15] and often to be placed away from the Fermi levels[16,17]. Consequently, tuning the filling factors of electronic flat bands or van Hove singularities, essential for accessing novel quantum phases, is challenging in bulk materials using standard techniques like gating or chemical doping.

Two-dimensional (2D) kagome materials offer promising solutions to these challenges. Inherently stable without capping layers, they effectively suppress couplings between kagome bands and bulk states. Their reduced thickness also lowers carrier densities, facilitating the experimental identification and manipulation of kagome bands. However, exfoliating 2D kagome materials from their bulk counterparts is difficult due to their non-van-der-Waals (vdW) interlayer interactions[12,13,18]. Even for recently reported vdW kagome materials like $Nb_3Cl_8$, successful exfoliation of its monolayer remains elusive[2,19]. Twisted moiré bilayers can host electronic kagome lattices[20–24] but introduce complications such as structural reconstructions[25–27], and issues on stability and scalability[28–32]. While monolayer kagome materials show potential[1,2,33–35], achieving and maintaining stable structures with desirable properties remains difficult. Progress has been made in constructing kagome monolayers on metal surfaces using organic molecules or inorganic clusters[36,37], but the resulting kagome-like band structures often arise from interactions with substrates[37–40] rather than intrinsic kagome bands. These issues persist across other approaches, like surface adatoms[41–43], intercalation[44,45], hybridization[46], and molecular manipulation[47]. Therefore, innovative strategies are necessitated to build pristine 2D kagome lattices[44].

Monolayer $MoTe_2$ and its analogues have been extensively studied for the formation of mirror twin boundaries (MTBs)[48–51] and resulting polymorphic phases[51,52]. These MTBs can form uniformly sized and ordered triangular loops[33,51,52], recently used to construct $Mo_5Te_8$[33] kagome monolayers. Here, we predicted a series of $MoTe_{2-x}$ kagome monolayers with varying MTB loop sizes and arrangements using density functional theory calculations. We established the relationship between MTB loop configurations and their resulting electronic structures near the Fermi level. Among ten configurations, we observed topological flat bands and Dirac states within the bandgap of the parent $MoTe_2$ monolayer, resulting in gapped (metallic) $MoTe_{2-x}$ monolayers. These states lead to ferromagnetism and potentially quantum spin Hall or correlated insulators. This systematically exploration of structural modifications and their effects on the electronic, magnetic, and topological properties paves the way for designing kagome materials with tailored properties for advanced applications.

**Results**

Monolayer $MoTe_2$ exhibits polymorphic phases. The most stable H-phase features a hexagonal atomic structure where Mo atoms are sandwiched between two layers of Te atoms, using the Mo layer as a mirror plane. Removal of some Te atoms from the H-phase monolayer leads to $MoTe_{2-x}$ monolayers with MTBs, which form distinct triangular loops (highlighted in light orange) essential for constructing MTB superlattices. As shown in Fig. 1a and Supplementary Fig. S1a, these loops, varying in size, are the building blocks of $MoTe_{2-x}$ monolayers. The size of an MTB loop is defined by the number (N) of $Te_2$ dimers it contains (denoted $Te_2$-T, balls in light pink). A triangular loop contains one $Te_2$ dimer is labeled as N1, three $Te_2$ dimers as N3, and six $Te_2$ dimers as N6, following the sequence of triangular numbers {1, 3, 6, 10, 15 …}.

Uniformly sized MTB triangles can be arranged in various configurations, such as vertex-to-vertex or vertex-to-edge. We classify these configurations according to the areas enclosed by the vertices of three MTB triangles. As shown in Fig. 1b, the vertices of three N6 MTB triangular loops form a small triangular area, marked with red dashed triangles, which contains no $Te_2$ dimers and is designated as T0, leading to the N6T0 configuration (Fig. 1b). In Figure 1c, a slight anticlockwise rotation of the N6 loops surrounds an area containing one $Te_2$ dimer (denoted $Te_2$-V, in red), leading to the N6T1 configuration. Further rotation closes three $Te_2$ dimers, forming the N6T3 configuration (Fig. 1d). The remaining areas are enclosed by the edges of MTB loops, in which the $Te_2$ dimers are termed $Te_2$-E (green balls in Fig. 1b to 1d). The "T" value in a monolayer reflects the twist (arrangement) of MTB loops. It equals to the number of $Te_2$-V dimers and is in the sequence of triangular numbers {0, 1, 3, 6, 10, 15…}. The T value must not exceed half of the N value except N1T1; for instance, N10T6 is meaningless and is equivalent to N10T3.

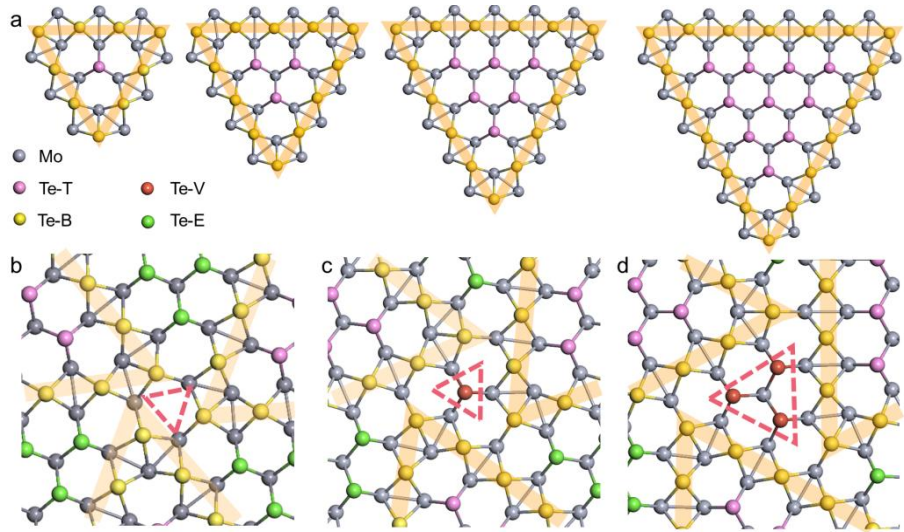

**Figure 1. Size and arrangement of triangular MTB loops in monolayer MoTe$_{2-x}$.** a) Atomic models of triangular MTB loops with N=1, 3, 6, and 10 (from left to right). Light orange triangles highlight the MTBs. Pink, yellow, red and green balls represent $Te_2$ dimers located inside the triangular MTB loops ($Te_2$-T), on the boundaries ($Te_2$-B), among the vertices ($Te_2$-V), and surrounded among the edges ($Te_2$-E), respectively. Slate blue balls represent Mo atoms. b-d) Atomic models of the T0 (b), T1 (c) and T3 (d) arrangements in monolayer MTB monolayers. Red dashed triangles outline the triangles surrounded by MTB vertices.

We considered ten monolayers by varying the sizes (N1, N3, N6, N10, and N15) and arrangements (T0, T1, and T3) of the triangular MTB loops (Supplementary Fig. S2). The supercell sizes of these monolayers range from 12.66 to 25.89 Å. The atomic lattice of these MTB monolayers belongs to two space groups (SG): P-62m (No. 189) and P-6 (No. 174). Specifically, the N1T1 and N6T3 monolayers fall under SG 189, exhibiting the $D_{3h}$ point group, which is the direct product of the $D_3$ point group and a mirror symmetry operation $\sigma_h$. Symmetry analysis of their atomic structures (Fig. S2) suggests these monolayers may form kagome-based lattices, marked with red triangles. The rest belong to SG 174, associated with a little point group of $C_{3h}$. Unlike SG 189 monolayers, these monolayers break the mirror symmetry perpendicular to the z-axis, resulting in a "breathing" distortion.

To establish the relationship between the atomic structures and the electronic properties in these nine $MoTe_{2-x}$ monolayers, we focused on analyzing the specific electronic states and band features of the T0 and T1 series, with N = 1, 3, 6,10 and15. Due to the complexity of kagome-related bands in the T3 and T6 series, it is difficult to clearly distinguish the occupied states, so these will not be discussed in detail here. For simplicity, spin-polarization was excluded to from the analysis so far.

Figure 2a presents the electronic bandstructure of the smallest T1 monolayer, N1T1 ($Mo_5Te_8$), which forms a coloring triangular (CT), a kagome variant, monolayer. Two kagome band sets, KBS1 (red) and KBS2 (blue), are separated by a ~0.1 eV bandgap around the Fermi level. These sets exhibit opposite signs for in-plane hopping parameters, with the valence band maximum (VBM) and conduction band minimum (CBM) positioned at the two flat bands, respectively, suggesting potential for an exciton insulator[53]. To verify the topological properties, spin-orbit coupling (SOC) was considered, opening a gap between the flat band and a Dirac band of KBS1 in N1T1. Our calculations yielded $Z_2$ invariants of 0, 1, and 1 for the three bands of set KBS1 with energies ordered from high to low, indicating a topological flat band. Due to the complex band-crossings in set KBS2, their topological properties are not further discussed in N1T1 and other monolayers.

The KBS1 set is primarily composed of the out-of-plane Mo-$d_{z^2}$ orbitals, which hybridize with the $d_{xy}$ and $d_{x^2-y^2}$ orbitals at the vertices of MTB loops (Fig. 2b), while the KBS2 set mainly arises from Mo-$d_{xz}$ and $d_{yz}$ states (Fig. 2c). Due to these distinct orbital characteristics, the size of the MTB loops (N value) has minimal impact on set KBS1, but as the MTB loops decrease in size, the in-plane confinement weakens, lowering the relative energy of the KBS2 bands compared to those of KBS1. This trend is confirmed in Figs. 2d to 2f, which show the band structures for the remaining T1 series as the N value increases. The energy position of the KBS1 set (red) is, relative to the vacuum level, nearly unchanged from N1 to N10, while that of the KBS2 set (blue) shifts downward, leading to partial superposition of the two sets in N6T1 and N10T1. Notably, significant electron hopping among the out-of-plane $d_{z^2}$ orbitals occurs in N1T1, where the MTB loop size is the smallest, but as the loop size increases, electron localization flattens the bands in KBS1.

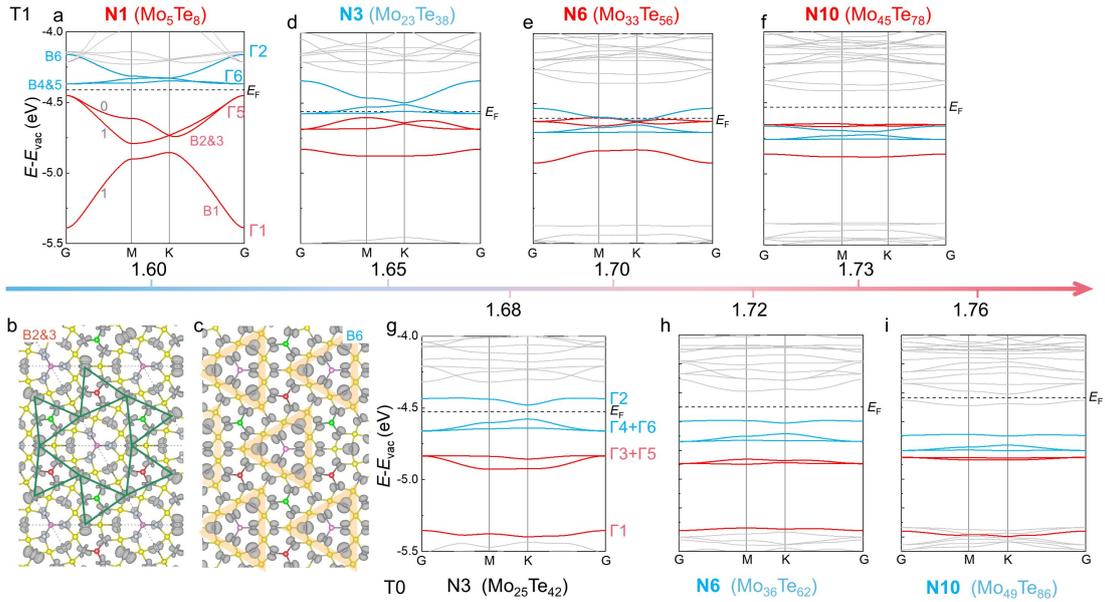

**Figure 2. Bandstructures of MoTe$_{2-x}$ monolayers with varying sizes and arrangements of MTB loops.** a) Bandstructure of N1T1, in which B1 to B6 represent six colored bands in increscent order of energy. Z$_2$ topological invariants of B1, B2 and B3 are marked with gray numbers in (a). b-c) Visualized wavefunction norms of bands B2 & B3 (b) and B6 (c) at the Γ point. The isosurface contours are 3×10$^{-3}$ e/Bohr$^3$. Dark green solid lines connect the Mo $d_{z^2}$ orbitals (hybridizing with $d_{xy}$ and $d_{x^2-y^2}$ orbitals), highlighting a coloring triangular lattice in (b). Light-orange shadowed triangles indicate the MTB loops in (c). d-f) Bandstructures of non-magnetic NxT1 (N=3, 6, 10) monolayers. h-i) Non-magnetic bandstructures of NxT0 (N=1, 3, 6,10) monolayers. The blue-red arrow in the middle of the figure indicates an increasing Te:Mo ratio $R_{\text{Te:Mo}}$ from 1.60 to 1.76, and specific values were written on both sides of the arrow. Specific sizes of the MTB loops and exact Mo and Te ratios are also provided, in which the red (blue) color indicates the most (comparable) stable monolayer, as elucidated later.

Compared to the 1:2 ratio in MoTe$_2$, the T1 series progressively transitions from having six fewer Te atoms in N1T1 (Mo$_5$Te$_8$), to eight fewer in N3T1 (Mo$_{23}$Te$_{38}$), ten fewer in N6T1 (Mo$_{33}$Te$_{56}$), and twelve fewer in N10T1 (Mo$_{45}$Te$_{78}$), which raises the Fermi level to higher energies. By adjusting the loop size, the occupancy of the six kagome bands from sets KBS1 and KBS2 can be precisely tuned, allowing control over the Fermi level. In N1T1 (Fig. 2a), the entire KBS1 set is occupied, and the flat band of KBS2 becomes pinned at the Fermi level, resulting in a metallic state with a high density of states in spin-degenerated bandstructure of N3T1 (Fig. 2d). One additional band from KBS2 is filled in N6T1 (Fig. 2e), the downward-shifted KBS2 bands partially overlap with the nearly unshifted KBS1 bands, leading to three narrowly distributed bands from the two sets crossing the Fermi level. These three bands, again, lead to a metallic state with a high density of states at the Fermi level in the spin-degenerated bandstructure. In the N10T1 configuration (Fig. 2f), both KBS1 and KBS2 sets are fully occupied, forming a band insulator state with flat bands near the band edge.

The T0 series exhibits bandstructure trends similar to the T1 series, with the KBS1 bands remaining nearly unchanged and KBS2 bands progressively lowering in energy. In the T0 series (Fig. 2g to 2i), the KBS2 Dirac bands open an approximately 0.1 eV bandgap at the K point for N3T0, decreasing to 0.06 eV for N10T0. As the MTB loop size increases, the bandwidth and the relative energy of the KBS2 bands decrease, while those for the KBS1 remains stable, which results in three types of bandgaps. As shown in Fig. 4g, the bandgap is at a non-degenerate Dirac cone (N3T0), which could be a quantum spin Hall insulator[54,55] if the two Dirac bands are topological non-trivial. For N6T0, the bandgap sits between two KBS sets (Fig. 2h). One more band is occupied for N10T0 and the bandgap resides across another non-degenerate Dirac cone (Fig. 2i). This tunability allows the Fermi level to pass through a split Dirac cone within a KBS set (Fig. 2g) or to lie above both KBS sets (Fig. 2h and 2i).

At the boundaries, Mo atoms exhibit two bonding types. They bond with three surrounding $Te_2$ dimers, including two $Te_2$-B and one $Te_2$-V dimer, and also share charges with neighboring Mo atoms. Specifically, Mo atoms at vertices (Mo-V) form bonds with two neighboring Mo atoms (Mo-VE), while those at edges (Mo-E) form only one Mo-Mo bond. In the T0 configuration, Te atoms at the MTB vertices bond to Mo-VE atoms, whereas in T1, T3, and T6, they bond with Mo-E atoms. This difference results in T0 consistently occupying one more KBS band than T1, T3, and T6 for the same triangular MTB loop size (N).

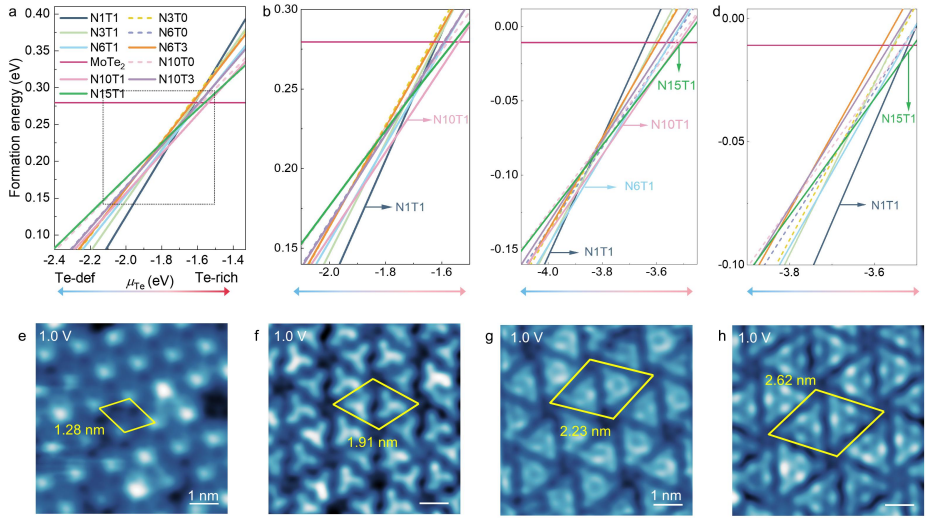

**Figure 3. Formation energies and STM topographic images of different monolayers**. a) Formation energies of the ten monolayers from Te-deficient to -rich conditions. b) The Magnified plot of the dashed box area in (a). The on-site U value is taken as 1.5 eV (a-b) and 0 eV (c) respectively. In panel (d), each monolayer is considered using its specified $U$ value. e-h) STM topography images of N1T1, N6T1, N10T1 and N15T1, with the bias voltage of 1.0 V. The red rhombuses indicate the supercell of the lattices, with lattice constants shown in red digits.

Figure 3a shows the formation energies of the nine configurations as a function of the chemical potential of Te, with Figs. 3b to 3d zooming in on their relative stability under Te-deficient conditions, from which we identify three most- and three comparable-stable monolayers. We also considered different on-site Coulomb energies ($U$) in our analysis, they are 1.5 (Fig. 3a and 3b) and 0.0 (Fig. 3c). In Fig. 3d, we compared the formation energy of each monolayer using its specific $U$ value. For N1T1, the Coulomb repulsion is stronger due to the smaller MTB size, with $U = 1.5$ eV aligning with the experimental gap. As the MTB size increases, $U$ is adjusted to 1.0, 0.5, and 0.0 eV for N3, N6, and N10 monolayers, respectively. These plots indicate that N1T1 ($Mo_5Te_8$), N6T1 ($Mo_{33}Te_{56}$), and N10T1 ($Mo_{45}Te_{78}$) are the most stable monolayers within specific Te chemical potential ranges under Te-deficient conditions. Given the most stable monolayers are all in the T1 series, we further considered N15T1 and found it has the lowest formation energy at certain Te chemical potentials. Notably, all four predicted stable T1 monolayers are experimentally realized, as shown in Fig. 3e to 3h. Additionally, we found that N3T1, N6T0, and N10T0 have comparable stability to the most stable monolayers across varying Te concentrations. These six monolayers are the focus of our discussion, as specific experimental conditions may alter the relative stability of $MoTe_{2-x}$, potentially promoting the growth of less stable monolayer predicted in Fig. 3.

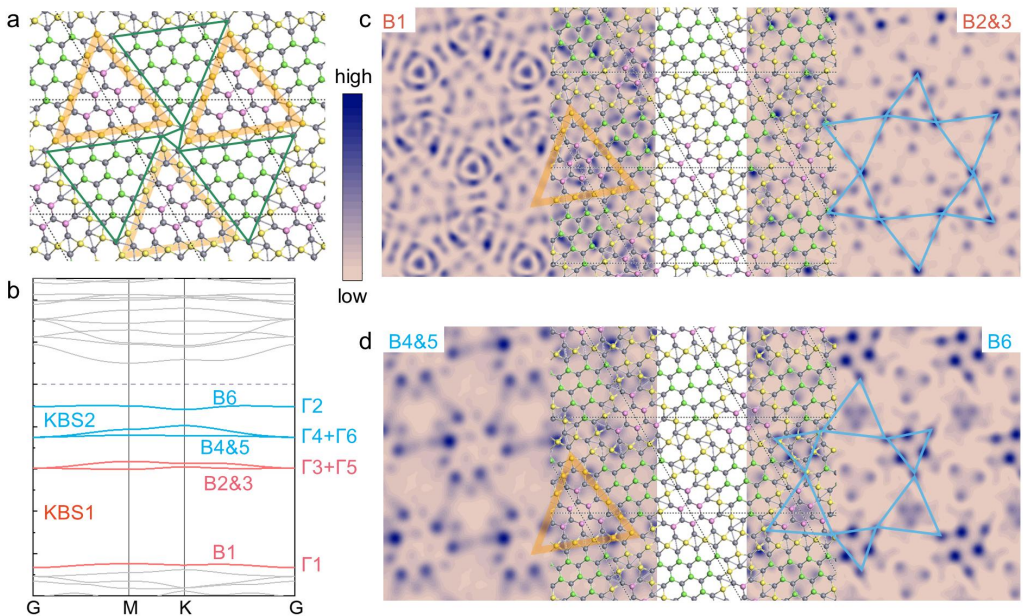

**Figure 4. Electronic structures of the N6T0 monolayer.** a) Atomic structure of the N6T0 monolayer, the color scheme follows that in Fig. 1 and Fig. 2b. b) Electronic band structures of N6T0. B1 to B6 represent six bands, belonging to kagome band sets KBS1 (in red) and KBS2 (in blue), while their irreducible representations are shown in the same color code. c-d) Simulated d$I$/d$V$ STS mapping images of B1 to B6 covered by the atomic structure diagrams. Light blue solid lines outline two distorted breaking kagome lattice formed by particular electronic states.

For the gapped monolayers, N6T0 (Fig. 4a) serves as a representative monolayer with SG 174 where the light pink lines connect Mo_V atoms, outlining a breathing coloring triangular lattice, a kagome variant. Figure 4b shows its electronic band structures near the Fermi level, indicating it is a non-magnetic, gapped, breathing kagome monolayer. Six electronic bands, observable within the plotted range, are labeled as B1 to B6 in ascending order of energy. We used the character table of the Γ/K little group to identify all irreducible representations (IRs) for the kagome-like bands (see in Table S1 and S2), using the Γ point (Γ1 to Γ6) for illustration. Two KB sets, comprising six bands in total, are connected through operator $\sigma_h$. Three bands of Γ1, Γ3, and Γ5 have a character of 1 under $\sigma_h$, corresponding to bands B1 to B3 (marked in red), forming a kagome band set, labeled KBS1. Meanwhile, bands Γ2, Γ4, and Γ6 have a character of -1 under $\sigma_h$, corresponding to bands B4 to B6 (marked in blue), forming the second KB set, labeled KBS2. Both sets are fully occupied and a nominal Dirac band acts as the VBM. Since the kagome lattices experimentally observed in STM imaging primarily arise from the electronic states of the surface Te atoms, we simulated d$I$/d$V$ STS mapping images for bands B1 to B6, as plotted in Figs. 4c and 4d. Due to the orbital hybridization between Mo and Te atoms, the dI/dV mapping images for both KBS1 and KBS2 appear as distinctly distorted kagome lattices, outlined with solid green lines.

Non-magnetic calculations (Fig. 2 and Fig. S3) reveal a high density of states at the Fermi level in N3T1, N6T1, and N6T3 monolayers, suggesting that they may undergo spontaneous spin-polarization to stabilize their Fermi surfaces. The calculated Stoner criteria for N3T1, N6T1, and N6T3 are 1.82, 2.79, and 1.32, respectively, suggesting Stoner ferromagnetism. Our DFT calculations confirm that the ferromagnetic (FM) states are more stable, with total energies at least 10 meV lower, than their non-magnetic (NM) counterparts. Figure 5a shows the spin-polarized band structure of N3T1, in which bands of the both KBSs appear around the Fermi level. The lower-energy set for the spin-up component (shown in orange) is fully occupied and the higher-energy set for the spin-down component (shown in green) is completely empty. Notably, two spin-down KBS1 bands cross the Fermi level. The spin-polarized DOS plot (Fig. 5b) shows a pronounced peak for the spin-up component residing between -0.20 and -0.10 eV, where the states are fully spin-polarized. Figure 5c depicts the spin density in the N3T1 monolayer, revealing that the majority spin (spin-up) is primarily concentrated on Mo atoms at the MTBs, while the minority spin (spin-down) is mainly distributed on Mo atoms at the vertices and within domain regions. This spatial distribution of spin densities aligns with the spatial contributions of the KBS1 and KBS2 sets. The spin-polarized band structures for N6T1 and N6T3 are presented in Fig. 2 of Ref. 56[56] and Fig. 5d, respectively, where the spin-down DOS peak crosses the Fermi level, resulting in a high DOS that may destabilize the Fermi surface for the N6T3 monolayer (Fig. 5e).

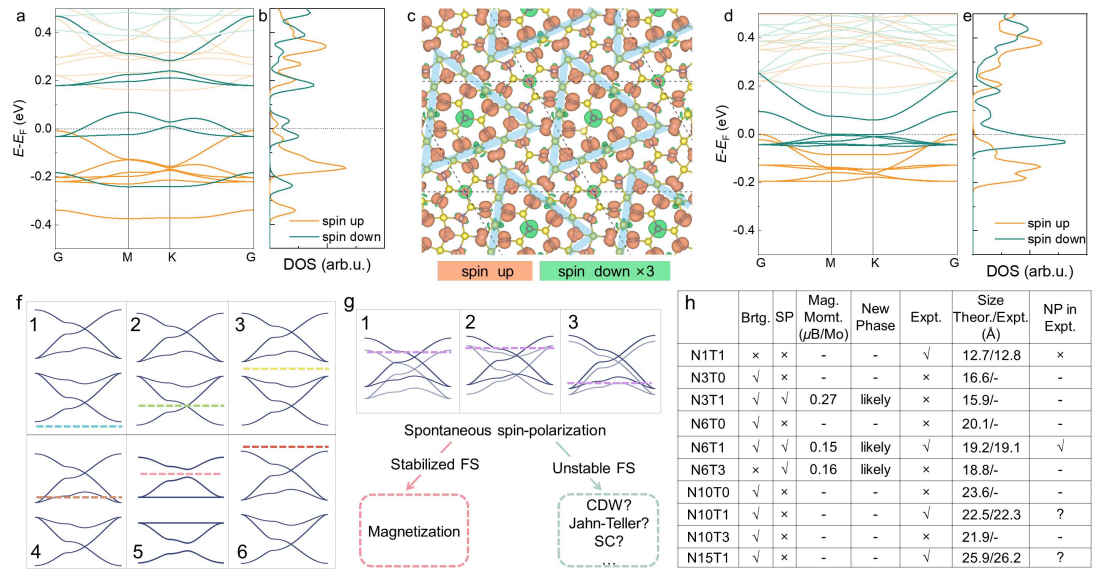

**Figure 5. Bandstructures and spin densities of ferromagnetic N3T1 and N6T3.** Spin-polarized band structures (a,d), corresponding density of states (DOS) (b,e) for N3T1 (a,b) and N6T3 (c,d). c) Spin density for N3T1. Light-bule shadowed triangles indicate the MTB loops. Orange and green colors represent spin-up and -down components. The isosurface contours are $2\times10^{-3}$ e/Bohr$^3$. f) Six cases of the position of the Fermi level for isolated KBSs. g) Three cases of partially overlapped KBSs, which can be classified into two categories based on the stability of the Fermi surface (FS), potentially leading to the formation of new phases (NPs). h) Different categories for all configurations distinguished by size, symmetry

and magnetization in theory (Theor.) and experiment (Expt.). The magnetic moment (Mag. Momt.) of each magnetic Mo atom is shown in (h).

**Summary and Discussion**

When the two KBSs are well isolated, the Fermi level can be positioned in various ways relative to their bands: below or above two Dirac bands (cases 1 and 6 in Fig. 5f, realized in N6T0 and N10T1), across a degenerate (case 2) or non-degenerate (case 5, N3T0) Dirac cone, between two flat bands (case 3, N1T1), or across a flat band (case 4, N3T1). In the case of partially overlapped KBSs (Fig. 5g), the Fermi level may intersect with two Dirac bands (case 1, N6T3), a flat and a Dirac band (case 2, N3T1 and N6T1), or two flat bands (case 3), resulting in a significantly high density of states at the Fermi level. Spontaneous spin-polarization could reduce this high DOS, leading to magnetization. However, in some cases, the DOS remains high even after spin splitting, potentially giving rise to correlated states, charge density waves (CDW), or other quantum phases. These phases can be further tuned by adjusting the filling factors using experimental techniques such as electrical gating.

All considered configurations could be classified according to their sizes (N1 to N15), atomic lattice symmetries [non-breathing (Non-Brtg.) and breathing (Brtg.)], spin polarization, and the DOS intensity at the Fermi level, as summarized in Fig. 5h. Except for N1T1, monolayers where the N value is not twice the T value form breathing kagome lattices. Additionally, these monolayers are categorized as either spin-polarized (SP) or spin-non-polarized (SNP), depending on the presence or absence of spontaneous magnetization. Notably, the T0 series is consistently spin-non-polarized, while the T1 and T3 series could be spin-polarized. All magnetic monolayers show high spin-polarized DOSs at their Fermi levels, suggesting additional quantum phases likely to be emerged in these monolayers.

In summary, we investigated the structural and electronic properties of MoTe$_{2-x}$ monolayers with various triangular MTB loop sizes and arrangements. We predicted that the stability of MTB monolayers can be tuned for enabling targeted structural growth by modulating the chemical potential of Te. All studied monolayers feature kagome-based structures with at least two sets of kagome bands near the Fermi level. Monolayers like N3T1, N6T1, N6T3, with high DOSs at the Fermi level, exhibit spontaneous magnetization due to Stoner ferromagnetism, highlighting their potential for verifying theoretical predictions on magnetism in kagome lattices. Additionally, we uncover two preliminary rules for designing the electronic structures of MTB monolayers, relevant with the bonding of Mo atoms and the size of MTB loops. The former rule ensures that the vertex-to-vertex MTB monolayer of a certain N (NxT0) always have one more occupied band than the vertex-to-edge ones of the same N value. The latter tells that the number of occupied kagome bands rises as the MTB loop size increases, shifting the Fermi level upwards. This tunability allows for precise optimization of the type of bandgaps and customization of electronic properties. Our work provides theoretical insights for further constructing and tuning monolayer kagome lattices, especially in H-phased transition metal dichalcogenides.

**Methods**
**DFT calculation:**

Our density functional theory (DFT) calculations were performed using the generalized gradient approximation (GGA) for the exchange correlation potential in the form of PerdewBurke–Ernzerhof (PBE)[57], the projector augmented wave method[58], and a plane-wave basis set as implemented in the Vienna ab-initio simulation package (VASP)[59]. Van der Waals interactions was performed at vdW-DF level for all calculations, with the optB86b functional for the exchange potential[60]. The energy cutoff for the plane-wave basis-sets was set to 500 eV for variable volume structural relaxation and electronic structure calculations. All atoms, lattice volumes, and shapes were allowed to relax until the residual force per atom was below 0.01 eV/Å. A vacuum layer exceeding 15 Å in thickness was employed to reduce imaging interactions between adjacent supercells. A Gamma-centered $k$-mesh of 5×5×1 and 3×3×1 was used to sample the first Brillouin zone of the unit cell for N1/N3 and N6/N10/N15, respectively. The Gaussian smearing method with a $\sigma$ value of 0.02 eV was applied for all calculations.

By using the Wannier90 package, we constructed the tight-binding model of $MoTe_2$ with Mo $4d$ and Te $5p$ orbitals based on the maximally localized Wannier functions method (MLWF)[61], performing on DFT calculations with SOC. We further calculated topological phases with $Z_2$ invariants using the WannierTools software package[62].

**Sample preparation:**

In this work, samples were grown on bilayer graphene (BLG), which was prepared by heating SiC (0001) in a home-built ultrahigh vacuum molecular beam epitaxy (MBE) system with a base pressure of approximately $1.0 \times 10^{-10}$ Torr. Te (99.999%) was evaporated from Knudsen cells, and high-purity Mo (99.95%) was evaporated from an e-beam evaporator, respectively. The flux ratios of Te/Mo were 30. The substrate was kept at approximately 200 °C. The deposition of $MoTe_2$ was followed by a 10-minute annealing at the growth temperature with the Te flux maintained. Subsequently, the substrate temperature was increased to 350 °C ~ 500 °C [56] and annealed for 1 hour to obtain the various ordered MTB superstructures.

**Scanning tunneling microscopy (STM) measurements:**

The STM measurements were carried out using a commercial Unisoku 1400 LT-STM system at 4.3 K (base pressure: $< 1 \times 10^{-10}$ Torr). Electrochemically etched W-tips were cleaned in situ with electron-beam bombardment and used in all measurements

**Acknowledgments**

We gratefully acknowledge the financial support from the Ministry of Science and Technology (MOST) of China (Grant No. 2023YFA1406500, Grant No. 2018YFE0202700), the National Natural Science Foundation of China (Grants No. 11974422 and 12104504), the Strategic Priority Research Program of Chinese Academy of Sciences (Grant No. XDB30000000), the Fundamental Research Funds for the Central Universities, and the Research Funds of Renmin University of China [Grants No. 22XNKJ30] (W.J.). J.D. was supported by the Outstanding Innovative Talents Cultivation Funded Programs 2023 of Renmin University of China. All calculations for this study were performed at the Physics Lab of High-Performance Computing (PLHPC) and the Public Computing Cloud (PCC) of Renmin University of China.